\documentclass[onecolumn]{IEEEtran}

\usepackage{amsmath}
\usepackage{amssymb}
\usepackage{graphicx}
\usepackage{psfrag}
\usepackage{subfigure}
\usepackage{url}
\usepackage{cite}

\newcommand{\C}{\ensuremath{\mathcal{C}}}
\newcommand{\Ct}[1]{\ensuremath{C^{#1}_t}}
\newcommand{\Cr}[1]{\ensuremath{C^{#1}_r}}
\newcommand{\Rt}[1]{\ensuremath{R^{#1}_t}}
\newcommand{\Rr}[1]{\ensuremath{R^{#1}_r}}
\newcommand{\Rpr}[1]{\ensuremath{R'_r}}
\newcommand{\Cn}{\ensuremath{C_n}}
\newcommand{\ns}{\!}

\newcommand{\E}{\ensuremath{\mathrm{E}}}

\newcommand{\fCt}[1]{\ensuremath{\overline{C}^{#1}_t}}
\newcommand{\fCr}[1]{\ensuremath{\overline{C}^{#1}_r}}
\newcommand{\fRt}[1]{\ensuremath{\overline{R}^{#1}_t}}
\newcommand{\fRr}[1]{\ensuremath{\overline{R}^{#1}_r}}
\newcommand{\fCn}{\ensuremath{\overline{C}_n}}

\providecommand{\abs}[1]{\lvert#1\rvert}
\providecommand{\asq}[1]{\abs{#1}^2}

\begin{document}

\title{The Impact of CSI and Power Allocation on Relay Channel Capacity and Cooperation Strategies}

\author{Chris~T.~K.~Ng,~\IEEEmembership{Member,~IEEE,}
        Andrea~J.~Goldsmith,~\IEEEmembership{Fellow,~IEEE}%
\thanks{This work was supported by the US Army under MURI award W911NF-05-1-0246, the ONR under award N00014-05-1-0168, and a grant from Intel.
Chris~T.~K.~Ng is supported by a Croucher Foundation Fellowship.
The material in this paper was presented in part at the IEEE International Symposium on Information Theory, Adelaide, Australia, September 2005,
and at the IEEE International Conference on Communications, Istanbul, Turkey, June 2006.}%
\thanks{Chris~T.~K.~Ng is with the Department of Electrical Engineering and Computer Science, Massachusetts Institute of Technology, Cambridge, MA 02139 USA (e-mail: ngctk@mit.edu).
Andrea~J.~Goldsmith is with the Department of Electrical Engineering, Stanford University, Stanford, CA 94305 USA (e-mail: andrea@wsl.stanford.edu).}%
}

\maketitle


\begin{abstract}

Capacity gains from transmitter and receiver cooperation are compared in a relay network where the cooperating nodes are close together.
Under quasi-static phase fading, when all nodes have equal average transmit power along with full channel state information (CSI), it is shown that transmitter cooperation outperforms receiver cooperation, whereas the opposite is true when power is optimally allocated among the cooperating nodes but only CSI at the receiver (CSIR) is available.
When the nodes have equal power with CSIR only, cooperative schemes are shown to offer no capacity improvement over non-cooperation under the same network power constraint.
When the system is under optimal power allocation with full CSI, the decode-and-forward transmitter cooperation rate is close to its cut-set capacity upper bound, and outperforms compress-and-forward receiver cooperation.
Under fast Rayleigh fading in the high SNR regime, similar conclusions follow.
Cooperative systems provide resilience to fading in channel magnitudes;
however, capacity becomes more sensitive to power allocation, and the cooperating nodes need to be closer together for the decode-and-forward scheme to be capacity-achieving.
Moreover, to realize capacity improvement, full CSI is necessary in transmitter cooperation, while in receiver cooperation optimal power allocation is essential.

\end{abstract}


\section{Introduction}
\label{sec:intro}

\PARstart{I}{n} ad hoc wireless networks, cooperation among nodes can be exploited to improve system performance.
In particular, cooperation among transmitters or receivers has been shown to improve capacity.
However, it is not clear if transmitter cooperation or receiver cooperation offers greater benefits in terms of capacity improvement.
The effectiveness of cooperation, moreover, may be dependent upon operating conditions such as the ease of acquiring channel state information (CSI) and the allocation of power among cooperating nodes.
In this paper, we evaluate the relative benefits of transmitter and receiver cooperation, and consider the cooperative capacity gain under different CSI and power allocation assumptions.

The benefits of node cooperation in wireless networks have been recently investigated by several authors.
The idea of cooperative diversity was pioneered in \cite{sendonaris03:coop1, sendonaris03:coop2}, where the transmitters cooperate by repeating detected symbols of other transmitters.
It was shown that cooperation can enlarge the achievable rate region when compared to the non-cooperative system.
In \cite{hunter02:coop_coding} the transmitters forward parity bits of the detected symbols, instead of the entire message, to achieve cooperation diversity.
Cooperative diversity and outage behavior was studied in \cite{laneman04:coop_diver}, where it was shown that orthogonal cooperative protocols can achieve full spatial diversity order.
Multiple-antenna systems and cooperative ad hoc networks were compared in
\cite{ng07:cap_tx_rx_coop, jindal04:cap_coop, ng04:txcoop_dcp_relay},
and the capacity gain from transmitter and receiver cooperation was characterized under different bandwidth allocation assumptions.
Information-theoretic achievable rate regions and bounds were derived in \cite{host-madsen06:coop_bounds, khojastepour04:coop_relay, host-madsen05:power_relay, host-madsen04:rx_coop, host-madsen03:coop_rate, kramer05:coop_cap_relay} for channels with transmitter and/or receiver cooperation.

In this work we consider the system model, first presented in
\cite{ng05:cap_txrx_coop, ng06:txrx_fading},
where a relay can be deployed either near the transmitter, or near the receiver.
Hence unlike previous works where the channel was assumed given, we treat the placement of the relay, and thus the resulting channel, as a design parameter.
Related works \cite{host-madsen05:power_relay, kramer05:coop_cap_relay, ahmed06:out_min_fading_relay, liang05:ortho_relay_alloc_cap} on relay channels have focused on characterizing capacity upper and lower bounds for a fixed network deployment.
Coding and bounding techniques examined in the previous works apply to both transmitter cooperation and receiver cooperation topologies; they do not resolve the relative performance between transmitter and receiver cooperation.
In this paper, we determine when is transmitter cooperation deployment better than that of receiver cooperation and vice versa, and we draw conclusions on effective relay deployment strategies under different operating scenarios.
Capacity improvement from cooperation is considered under system models of full CSI or CSI at the receiver (CSIR) only, with optimal or equal power allocation among the cooperating nodes.
To gain intuition on the relative benefits of transmitter and receiver cooperation, we first consider a simple model where the channels experience quasi-static phase fading with constant magnitudes.
Then we extend the analysis to consider channels under fast Rayleigh fading in the high signal-to-noise ratio (SNR) regime.

The rest of the paper is organized as follows.
Section~\ref{sec:sys_mod} presents the system model, the different CSI and power allocation assumptions, and the transmitter and receiver cooperation strategies.
Section~\ref{sec:qs_static_ch} evaluates the relative benefits of transmitter and receiver cooperation under quasi-static phase fading.
Fast Rayleigh fading is considered in Section~\ref{sec:ray_fading_ch}, in which
we examine the effects of fading on transmitter and receiver cooperation, and identify the necessary conditions for cooperative capacity gain.
Section~\ref{sec:conclu} concludes the paper.

\section{System Model}
\label{sec:sys_mod}

\subsection{Channel Model}

\begin{figure}
  \centerline{%
    \subfigure[Transmitter cooperation]{%
      \psfrag{e1}[][]{\small$h_1$}
      \psfrag{e2}[][]{\small$\sqrt{g}h_2$}
      \psfrag{e3}[][]{\small$h_3$}
      \includegraphics{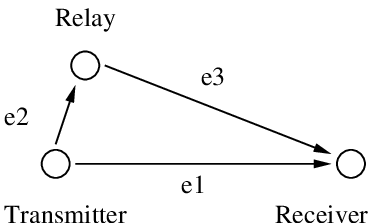}
      \label{fig:txcoop}
    }
    \hfil
    \subfigure[Receiver cooperation]{%
      \psfrag{e1}[][]{\small$h_1$}
      \psfrag{e2}[][]{\small$h_2$}
      \psfrag{e3}[][]{\small$\sqrt{g}h_3$}
      \includegraphics{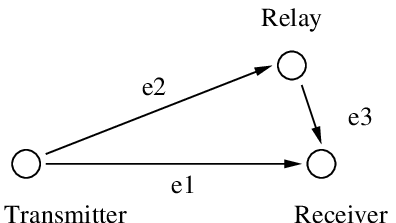}
      \label{fig:rxcoop}
    }
  }
  \caption{Cooperation system model.}
  \label{fig:coop}
\end{figure}

Consider a discrete-time frequency-flat single-antenna channel with additive white Gaussian noise (AWGN).
To exploit cooperation, a relay can be deployed either close to the transmitter to form a transmitter cluster, or close to the receiver to form a receiver cluster, as illustrated in Fig.~\ref{fig:coop}.
The transmitter cooperation relay network in Fig.~\ref{fig:txcoop} is described by
\begin{align}
  \label{eq:tx_relay}
  y_1 & = \sqrt{g}h_2 x + n_1, &
  y & = h_1 x + h_3 x_1 + n,
\end{align}
where $x,y,n,x_1,y_1,n_1,h_i\in\mathbb{C}$, $i=1,2,3$, $g\in\mathbb{R}_+$: $x$ is the signal sent by the transmitter, $y$ is the signal received by the receiver, $y_1,x_1$ are the received and transmitted signals of the relay, respectively, and $n,n_1$ are independent identically distributed (iid) zero-mean circularly symmetric complex Gaussian (ZMCSCG) noise with unit variance.
We assume information-theoretical full-duplex relaying; i.e., the relay can receive and transmit in the same band simultaneously \cite{cover79:cap_relay, kramer05:coop_cap_relay}.
Similarly, the receiver cooperation relay network in Fig.~\ref{fig:rxcoop} is given by
\begin{align}
  \label{eq:rx_relay}
  y_1 & = h_2x + n_1, &
  y & = h_1x + \sqrt{g}h_3x_1 + n.
\end{align}
We assume the average path loss power attenuation between the clusters is normalized to unity with $\E[\asq{h_i}]=1$, while within the cluster it is denoted by $g$.
As the cooperating nodes are assumed to be close to each other, the scenario of interest is when $g$ is large.

We consider two models of channel fading.
In Section~\ref{sec:qs_static_ch},
we focus on investigating the effects of node geometry and assume quasi-static phase fading \cite{kramer05:coop_cap_relay} channels with constant magnitudes:
$h_i = e^{j\theta_i}$, where $\theta_i\sim U[0,2\pi]$ is iid uniform, and $\theta_i$ remains unchanged after its realization.
Then in Section~\ref{sec:ray_fading_ch} we consider fast Rayleigh fading where the channel gain $h_i$ is iid ZMCSCG with unit variance; hence the corresponding power gain is iid exponential with unit mean: $\gamma_i\triangleq\asq{h_i}\sim\exp(1)$.

The output of the relay depends causally on its past inputs.
We assume an average network power constraint on the system:
$\E\bigl[\asq{x} + \asq{x_1}\bigr] \leq P$, where the expectation is taken over repeated channel uses.
The network power constraint model is applicable when the node configuration in the network is not fixed.
For example, in an ad hoc wireless network, a particular node might take on the role of the transmitter over some time period, but act as the relay over another time period.

\subsection{CSI and Power Allocation}

We compare the rate achieved by transmitter cooperation versus that by receiver cooperation under different operating conditions.
Under quasi-static phase fading in Section~\ref{sec:qs_static_ch},
we consider two models of CSI:
i)~every node has full CSI (i.e., the transmitter, relay and receiver all know $\theta_1,\theta_2,\theta_3$ and $g$); and ii)~the nodes have CSIR only
(i.e., the relay knows $\theta_2$, the receiver knows $\theta_1, \theta_3$, and the transmitter, relay and receiver all know $g$).
In both cases we assume all nodes have knowledge of the channel distribution information (CDI).
For the Rayleigh fading channels in Section~\ref{sec:ray_fading_ch},
we assume a fast-fading environment, where the receiver has CSI to perform coherent detection, but there is no fast feedback link to convey the CSI to the transmitter.
Hence the transmitter has only CDI, but no knowledge of the instantaneous channel realizations.
We assume the duration of the transmission codewords is long enough to undergo all channel realizations, and thus ergodic capacity is used to characterize the transmission rate of the channel.

Independent of the fading and CSI assumptions, we consider two models of power allocation:
i)~power is optimally allocated between the transmitter and the relay,
i.e., $\E\bigl[\asq{x}\bigr] \leq \alpha P$,
$\E\bigl[\asq{x_1}\bigr] \leq (1-\alpha) P$, where $\alpha \in [0,1]$
is a parameter to be optimized based on CDI and node geometry $g$;
ii)~the network is homogeneous and all nodes have equal average power constraints, i.e.,
$\E\bigl[\asq{x}\bigr]\leq P/2$ and $\E\bigl[\asq{x_1}\bigr]\leq P/2$.
Power allocation for AWGN relay channels with arbitrary channel gains was treated in \cite{host-madsen05:power_relay}, and for fading relay channels with CSI at the transmitter (CSIT) in \cite{ahmed06:out_min_fading_relay, liang05:ortho_relay_alloc_cap}; in this paper we consider only the case when the cooperating nodes form a cluster, and under fast Rayleigh fading we assume the transmitter is not able to adapt to the instantaneous channel realizations.

\subsection{Cooperation Strategies}

The three-terminal network shown in Fig.~\ref{fig:coop} is a relay channel \cite{meulen71:3t_comm_ch, cover79:cap_relay}, and its capacity is not known in general.
The cut-set bound described in \cite{cover79:cap_relay, cover91:eoit} provides a capacity upper bound. Achievable rates obtained by two coding strategies were also given in \cite{cover79:cap_relay}.
The first strategy \cite[Thm.~1]{cover79:cap_relay} has become known as (along with other slightly varied nomenclature) ``decode-and-forward'' \cite{laneman04:coop_diver, kramer05:coop_cap_relay, host-madsen06:coop_bounds}, and the second one \cite[Thm.~6]{cover79:cap_relay} is called ``compress-and-forward''
\cite{kramer05:coop_cap_relay, host-madsen05:power_relay, khojastepour04:coop_relay}.
In particular, it was shown in \cite{kramer05:coop_cap_relay} that decode-and-forward approaches capacity (and achieves capacity under certain conditions) when the relay is near the transmitter, whereas compress-and-forward is close to optimum when the relay is near the receiver.
Therefore, in our analysis decode-and-forward is used in transmitter cooperation, while compress-and-forward is used in receiver cooperation.

To evaluate the capacity gain from cooperation, the transmitter and receiver cooperation rates are compared to the non-cooperative capacity when the relay is not available, i.e., the capacity of a single-user channel under the same network power constraint $P$\@.
Therefore, under quasi-static phase fading in Section~\ref{sec:qs_static_ch}, the non-cooperative capacity $\Cn$ is given by $\C(1)$, where $\C(x) \triangleq \log(1+xP)$.
Whereas under fast Rayleigh fading in Section~\ref{sec:ray_fading_ch}, the non-cooperative ergodic capacity is given by $\fCn = \E\bigl[\C(\gamma_1)\bigr]$.
In this paper all logarithms are base~2.

\section{Quasi-Static Phase Fading}
\label{sec:qs_static_ch}

In this section we assume the channels in the relay network are under quasi-static phase fading, i.e., $h_i = e^{j\theta_i}$, $\theta_i\sim U[0,2\pi]$ iid, $i=1,2,3$.
We consider the cases when the nodes have full CSI or CSIR only, and the cases when the network is under optimal power allocation or equal power allocation.
Considering the combinations of the different CSI and power allocation models, Table~\ref{tab:cases} enumerates the four cases under which the relative benefits of transmitter and receiver cooperation are investigated.
In terms of notation, a capacity upper bound is denoted by $C$, whereas an achievable rate is denoted by $R$. The subscript $t$ represents transmitter cooperation, and $r$ represents receiver cooperation.
A superscript is used, when applicable, to denote the case under consideration; e.g., $\Ct{1}$ corresponds to the transmitter cut-set bound in Case~1: Optimal power allocation with full CSI\@.

\begin{table}
  \renewcommand{\arraystretch}{1.2}
  \centering
  \begin{tabular}{|c|l|}
    \hline
    \emph{Case} & \emph{Description}\\
    \hline\hline
    Case~1 & Optimal power allocation with full CSI\\
    \hline
    Case~2 & Equal power allocation with full CSI\\
    \hline
    Case~3 & Optimal power allocation with CSIR only\\
    \hline
    Case~4 & Equal power allocation with CSIR only\\
    \hline
  \end{tabular}
  \caption{Cooperation under different operating conditions}
  \label{tab:cases}
\end{table}

\subsection{Transmitter Cooperation}

Suppose that the transmitter is operating under an average power constraint $\alpha P$, $0\leq\alpha\leq 1$, and the relay under constraint $(1-\alpha)P$.
Then for the transmitter cooperation network depicted in Fig.~\ref{fig:txcoop}, the cut-set bound \cite{cover91:eoit} is
\begin{align}
  \label{eq:Ct}
    \Ct{} = \max_{\substack{0\leq\rho\leq 1}}
    \min\Bigl\{\C\Bigl(\alpha(g+1)(1-\rho^2)\Bigr),\,
    \C\Bigl(1+2\rho\sqrt{\alpha(1-\alpha)}\Bigr)\Bigr\},
\end{align}
where $\rho$ represents the correlation between the transmit signals of the transmitter and the relay. With optimal power allocation in Case~1 and Case~3, $\alpha$ is to be further optimized, whereas $\alpha=1/2$ in Case~2 and Case~4 under equal power allocation.

In the decode-and-forward transmitter cooperation strategy, transmission is done in blocks: the relay first fully decodes the transmitter's message in one block, then in the ensuing block the relay and the transmitter cooperatively send the message to the receiver.
The decode-and-forward transmitter cooperation rate is given in \cite{kramer05:coop_cap_relay}:
\begin{align}
  \label{eq:Rt}
  \begin{split}
    \Rt{} = \max_{\substack{0\leq\rho\leq 1}}
    \min\Bigl\{\C\Bigl(\alpha g(1-\rho^2)\Bigr),\,
    \C\Bigl(1+2\rho\sqrt{\alpha(1-\alpha)}\Bigr)\Bigr\},
  \end{split}
\end{align}
where $\rho$ and $\alpha$ carry similar interpretations as described above in (\ref{eq:Ct}).
Note that $\Rt{}\bigr\rvert_{g} = \Ct{}\bigr\rvert_{g-1}$ for $g\geq 1$, which can be used to aid the calculation of $\Rt{}$ in the subsequent sections.

\subsection{Receiver Cooperation}

For the receiver cooperation network shown in Fig.~\ref{fig:rxcoop}, the cut-set bound is
\begin{align}
  \label{eq:Cr}
    \Cr{} = & \max_{\substack{0\leq\rho\leq 1}}
    \min\Bigl\{\C\Bigl(2\alpha(1-\rho^2)\Bigr),\,
    \C\Bigl(\alpha+(1-\alpha)g
    +2\rho\sqrt{\alpha(1-\alpha)g}\Bigr)\Bigr\}.
\end{align}
In the compress-and-forward receiver cooperation strategy, the relay sends a compressed version of its observed signal to the receiver.
The compression is realized using Wyner-Ziv source coding \cite{wyner76:rate_dist_side_info}, which exploits the correlation between the received signal of the relay and that of the receiver.
The compress-and-forward receiver cooperation rate is given in \cite{kramer05:coop_cap_relay, host-madsen05:power_relay}:
\begin{align}
  \label{eq:Rr}
  \Rr{} = \C\Bigl(\textstyle\frac{\alpha(1-\alpha)g}
    {(1-\alpha)g+2\alpha+1/P}+\alpha\Bigr).
\end{align}
Likewise, in (\ref{eq:Cr}) and (\ref{eq:Rr}) $\alpha$ needs to be optimized in Case~1 and Case~3, and $\alpha=1/2$ in Case~2 and Case~4.

\subsection{Evaluation of Cooperation Rates}
\label{sec:coop_rates}

\subsubsection*{Case 1: Optimal power allocation with full CSI}
\label{sec:case1}

Consider the transmitter cooperation cut-set bound in (\ref{eq:Ct}).
Recognizing the first term inside $\min\{\cdot\}$ is a decreasing function of $\rho$, while the second one is an increasing one, the optimal $\rho^*$ can be found by equating the two terms (or maximizing
the lesser term if they do not equate).
Next the optimal $\alpha^*$ can be calculated by setting its derivative to zero.
The other upper bounds and achievable rates, unless otherwise noted, can be optimized using similar techniques; thus in the following sections they will be stated without repeating the analogous arguments.

The transmitter cooperation cut-set bound is found to be
\begin{align}
  \label{eq:Ct1}
  \Ct{1} = \C\bigl(\textstyle\frac{2(g+1)}{g+2}\bigr),
\end{align}
with $\rho^* = \sqrt{g/(g+4)}$, $\alpha^* = (g+4)/(2g+4)$.
The decode-and-forward transmitter cooperation rate is
\begin{align}
  \label{eq:Rt1}
  \Rt{1} = \Bigl\{\textstyle
  \C\bigl(\frac{2g}{g+1}\bigr) \text{ if } g\geq1,
  \quad \C(g) \text{ if } g<1 \Bigr.,
\end{align}
with $\rho^*=\sqrt{(g-1)/(g+3)}$, $\alpha^*=(g+3)/(2g+2)$ if $g\geq1$, and $\rho^*=0$, $\alpha^*=1$ otherwise.  It can be observed that the transmitter cooperation rate $\Rt{1}$ in (\ref{eq:Rt1}) is close to its upper bound $\Ct{1}$ in (\ref{eq:Ct1}) when $g\gg1$.

For receiver cooperation, the cut-set bound is given by
\begin{align}
  \label{eq:Cr1}
  \Cr{1} = \C\bigl(\textstyle\frac{2(g+1)}{g+2}\bigr),
\end{align}
with $\rho^* = 1/\sqrt{g^2+2g+2}$, $\alpha^*=(g^2+2g+2)/(g^2+3g+2)$.
For the compress-and-forward receiver cooperation rate $\Rr{1}$, the expression of the optimal value $\alpha^*$  in (\ref{eq:Rr}) does not have a simple form that permits straightforward comparison of $\Rr{1}$ with the other upper bounds and achievable rates.
Instead we consider an upper bound to $\Rr{1}$, which has a simpler expression.
The upper bound $\Rpr\ $ is obtained by omitting the term $1/P$ in the denominator in (\ref{eq:Rr}) as follows:
\begin{align}
  \label{eq:Rr1}
  \Rr{1} & = {\max_{0\leq\alpha\leq1}}
  \C\Bigl(\textstyle
    \frac{\alpha(1-\alpha)g}{(1-\alpha)g+2\alpha+1/P}+\alpha\Bigr)\\
  \label{eq:Rr1_def}
  & < {\max_{0\leq\alpha\leq1}}\C\Bigl(\textstyle
  \frac{\alpha(1-\alpha)g}{(1-\alpha)g+2\alpha}+\alpha\Bigr)
  \triangleq \Rpr\ .
\end{align}
Since the term $(1-\alpha)g+2\alpha$ in the denominator in (\ref{eq:Rr1}) ranges between 2 and $g$, the upper bound $\Rpr\ $ in (\ref{eq:Rr1_def}) is tight when $g>2$ and $P\gg1$. Specifically, for $g>2$, the receiver cooperation rate upper bound is found to be
\begin{align}
  \label{eq:Rpr}
  \Rpr\ = \C\Bigl(\textstyle
  \frac{2g(\sqrt{g-1}-1)(g-1-\sqrt{g-1})}{\sqrt{g-1}(g-2)^2}\Bigr),
\end{align}
with the upper bound's optimal $\alpha^*=\frac{g(g-1-\sqrt{g-1})}{g^2-3g+2}$.

Note that the transmitter and receiver cut-set bounds $\Ct{1}$ and $\Cr{1}$ are identical.
However, for $\infty>g>1$, the decode-and-forward transmitter cooperation rate $\Rt{1}$
outperforms the compress-and-forward receiver cooperation upper bound $\Rpr\ $.
Moreover, the decode-and-forward rate is close to the cut-set bounds when $g\geq2$; therefore, transmitter cooperation is the preferable strategy when the system is under optimal power allocation with full CSI\@.

\begin{figure}
  \centering
  \includegraphics*[width=8cm]{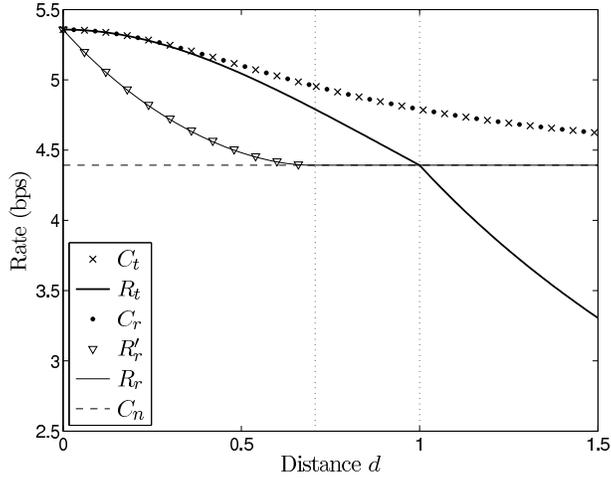}
  \caption{Cooperation cut-set bounds and achievable rates in Case~1.}
  \label{fig:rates_1}
\end{figure}

Numerical examples of the upper bounds and achievable rates, together with the non-cooperation capacity $\Cn$, are shown in Fig.~\ref{fig:rates_1}.
Note that as decode-and-forward requires that the relay fully decodes the transmitter's message, the transmitter cooperate rate can be lower than the non-cooperation transmission rate when the relay and transmitter are far apart.
On the other hand, compress-and-forward, having no requirement on decoding, always equals or outperforms non-cooperation under optimal power allocation.
In all plots of the numerical results, we assume the channel has unit bandwidth, the system has an average
network power constraint $P$ = 20, and $d$ is the distance between the relay and its cooperating node.
We assume a path-loss power attenuation exponent of 2, and hence $g=1/d^2$.
The vertical dotted lines mark $d=1/\sqrt{2}$ and $d=1$, which correspond to $g=2$ and $g=1$, respectively.
We are interested in the capacity improvement when the cooperating nodes are close together, and $d<1/\sqrt{2}$ (or $g>2$) is the region of our main focus.

\subsubsection*{Case 2: Equal power allocation with full CSI}
\label{sec:case2}

With equal power allocation, both the transmitter and the relay are under an average power constraint of $P/2$, and so $\alpha$ is set to $1/2$.
For transmitter cooperation, the cut-set capacity upper bound is found to be
\begin{align}
  \label{eq:Ct2}
  \Ct{2} = \Bigl\{\textstyle
    \C\bigl(\frac{2g}{g+1}\bigr) \text{ if } g\geq1,
    \quad \C\bigl(\frac{1+g}{2}\bigr) \text{ if } g<1 \Bigr.,
\end{align}
with $\rho^*=(g-1)/(g+1)$ if $g\geq1$, and $\rho^*=0$ otherwise.
Incidentally, the bound $\Ct{2}$ in (\ref{eq:Ct2}) coincides with the transmitter cooperation rate $\Rt{1}$ in (\ref{eq:Rt1}) obtained in Case~1 for $g\geq1$.
Next, the decode-and-forward transmitter cooperation rate is given by
\begin{align}
  \label{eq:Rt2}
  \Rt{2} = \Bigl\{\textstyle
  \C\bigl(\frac{2(g-1)}{g}\bigr) \text{ if } g\geq2,
  \quad \C\bigl(\frac{g}{2}\bigr) \text{ if } g<2 \Bigr.,
\end{align}
with $\rho^*=(g-2)/g$ if $g\geq2$, and $\rho^*=0$ otherwise.
Similar to Case~1, the transmitter cooperation rate $\Rt{2}$ in (\ref{eq:Rt2}) is close to its upper bound $\Ct{2}$ in (\ref{eq:Ct2}) when $g\gg1$.

For receiver cooperation, the corresponding cut-set bound resolves to
\begin{align}
  \label{eq:Cr2}
  \Cr{2} = \Bigl\{\textstyle
  \C(1) \text{ if } g\geq1,
  \quad \C\Bigl(\frac{1+\sqrt{g(2-g)}}{2}\Bigr) \text{ if } g<1
  \Bigr.,
\end{align}
with $\rho^* = 0$ for $g\geq1$, and $\rho^*=(\sqrt{2-g}-\sqrt{g})/2$ otherwise.
Lastly, the compress-and-forward receiver cooperation rate is
\begin{align}
  \label{eq:Rr2}
  \Rr{2} = \C\bigl(\textstyle\frac{g}{2(g+2+2/P)}+\frac{1}{2}\bigr).
\end{align}

\begin{figure}
  \centering
  \includegraphics*[width=8cm]{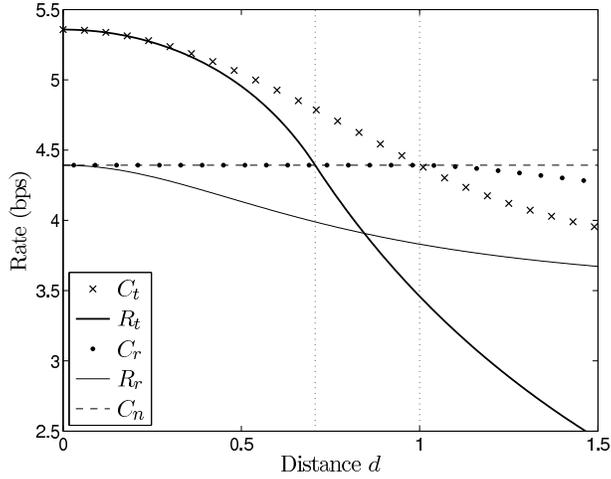}
  \caption{Cooperation cut-set bounds and achievable rates in Case~2.}
  \label{fig:rates_2}
\end{figure}

It can be observed that if the cooperating nodes are close together such that $g>2$, the transmitter cooperation rate $\Rt{2}$ is strictly higher than the receiver cooperation cut-set bound $\Cr{2}$;
therefore, transmitter cooperation conclusively outperforms receiver cooperation when the system is under equal power allocation with full CSI\@.
Fig.~\ref{fig:rates_2} illustrates the transmitter and receiver cooperation upper bounds and achievable rates.

\subsubsection*{Case 3: Optimal power allocation with CSIR}
\label{sec:case3}

When CSIT is not available, it was derived in \cite{kramer05:coop_cap_relay, host-madsen05:power_relay} that it is
optimal to set $\rho=0$ in the cut-set bounds (\ref{eq:Ct}), (\ref{eq:Cr}), and the decode-and-forward transmitter cooperation rate (\ref{eq:Rt}).
Intuitively, with only CSIR, the relay and the transmitter, being unable to realize the gain from coherent combining, resort to sending uncorrelated signals.
The receiver cooperation strategy of compress-and-forward, on the other hand, does not require phase information at the transmitter \cite{kramer05:coop_cap_relay}, and thus the receiver cooperation rate is still given by (\ref{eq:Rr}) with the power allocation parameter $\alpha$ optimally chosen.

Under the transmitter cooperation configuration, the cut-set bound is found to be
\begin{align}
  \label{eq:Ct3}
  \Ct{3} = \C(1),
\end{align}
where $\alpha^*$ is any value in the range $[1/(g+1), 1]$.
When the relay is close to the transmitter ($g\geq1$), the decode-and-forward strategy is capacity achieving, as reported in \cite{kramer05:coop_cap_relay}.
Specifically, the transmitter cooperation rate is given by
\begin{align}
  \label{eq:Rt3}
  \Rt{3} = \Bigl\{\textstyle
  \C(1) \text{ if } g\geq1,
  \quad \C(g) \text{ if } g<1 \Bigr.,
\end{align}
where $\alpha^*$ is any value in the range $[1/g, 1]$ if $g\geq1$, and
$\alpha^*=1$ otherwise.

For receiver cooperation, the cut-set bound is
\begin{align}
  \label{eq:Cr3}
  \Cr{3} = \Bigl\{\textstyle
  \C\bigl(\frac{2g}{g+1}\bigr) \text{ if } g\geq1,
  \quad \C(1) \text{ if } g<1 \Bigr.,
\end{align}
where $\alpha^*=g/(g+1)$ if $g>1$, $\alpha^*=1$ if $g<1$, and $\alpha^*$ is any value in the range $[g/(g+1), 1]$ if $g=1$.
Since compress-and-forward does not make use of phase information at the transmitter, the receiver cooperation rate is the same as (\ref{eq:Rr1}) given in Case~1: $\Rr{3}=\Rr{1}$.
Note that the argument inside $\C(\cdot)$ in (\ref{eq:Rr1}) is 1 when $\alpha=1$, and hence $\Rr{3}\geq\C(1)$.

\begin{figure}
  \centering
  \includegraphics*[width=8cm]{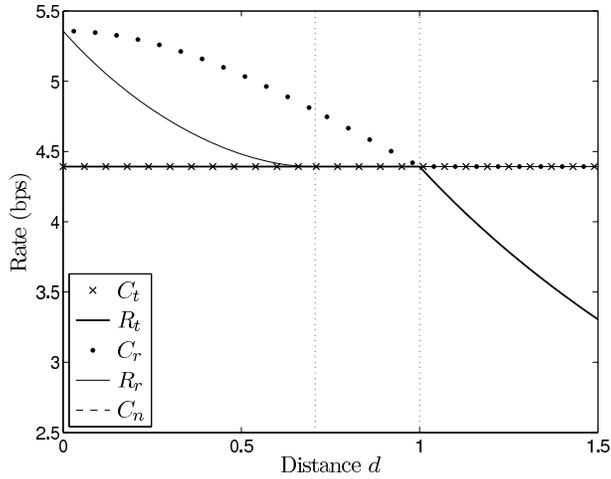}
  \caption{Cooperation cut-set bounds and achievable rates in Case~3.}
  \label{fig:rates_3}
\end{figure}

In contrast to Case~2, the receiver cooperation rate $\Rr{3}$ equals or outperforms the transmitter cooperation cut-set bound $\Ct{3}$; consequently receiver cooperation is the superior strategy when the system is under optimal power allocation with CSIR only.
Numerical examples of the upper bounds and achievable rates are shown in Fig.~\ref{fig:rates_3}.

\subsubsection*{Case 4: Equal power allocation with CSIR}
\label{sec:case4}

With equal power allocation, $\alpha$ is set to $1/2$.
With only CSIR, similar to Case~3, $\rho=0$ is optimal for the cut-set bounds and decode-and-forward rate. Therefore, in this case no optimization is necessary, and the bounds and achievable rates can be readily evaluated.

For transmitter cooperation, the cut-set bound and the decode-and-forward rate, respectively, are
\begin{align}
  \label{eq:Ct4}
  \Ct{4} & = \Bigl\{\textstyle
  \C(1) \text{ if } g\geq1,
  \quad \C\bigl(\frac{1+g}{2}\bigr)  \text{ if } g<1 \Bigr.,\\
  \label{eq:Rt4}
  \Rt{4} & = \Bigl\{\textstyle
  \C(1) \text{ if } g\geq2,
  \quad \C\bigl(\frac{g}{2}\bigr) \text{ if } g<2 \Bigr..
\end{align}
For receiver cooperation, the cut-set bound is
\begin{align}
  \label{eq:Cr4}
  \Cr{4} = \Bigl\{\textstyle
  \C(1) \text{ if } g\geq1,
  \quad\C\bigl(\frac{1+g}{2}\bigr)\text{ if } g<1\Bigr.,
\end{align}
and the compress-and-forward rate is the same as (\ref{eq:Rr2}) in Case~2: $\Rr{4}=\Rr{2}$.

\begin{figure}
  \centering
  \includegraphics*[width=8cm]{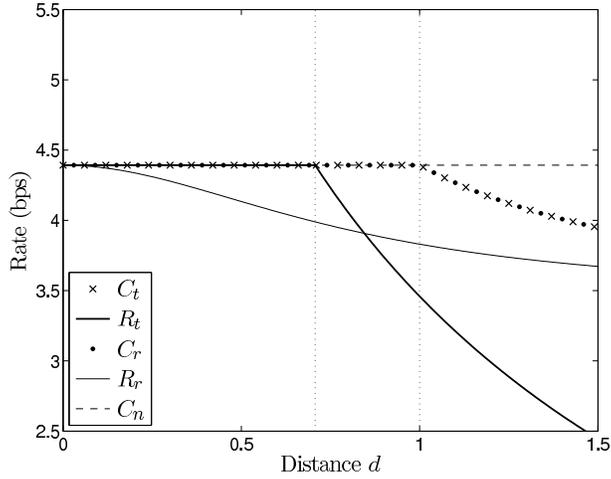}
  \caption{Cooperation cut-set bounds and achievable rates in Case~4.}
  \label{fig:rates_4}
\end{figure}

Parallel to Case~1, the transmitter and receiver cooperation cut-set bounds $\Ct{4}$ and $\Cr{4}$ are identical. Note that the non-cooperative capacity $\Cn$ meets the cut-set bounds when $g\geq1$, and even beats the bounds when $g<1$.
Hence it can be concluded cooperation offers no capacity improvement when the system is under
equal power allocation with CSIR only.
Numerical examples of the upper bounds and achievable rates are plotted in Fig.~\ref{fig:rates_4}.

\subsection{Effects of CSI and Power Allocation}

In the previous sections, for each given operating condition we derive the most advantageous cooperation strategy.
However, the available mode of cooperation is sometimes dictated by practical system constraints.
For instance, in a wireless sensor network collecting measurements for a single remote base station, only transmitter cooperation is possible.
In this section, for a given transmitter or receiver cluster, the trade-off between cooperation capacity gain and
the requirements on CSI and power allocation is examined.

The upper bounds and achievable rates from the previous sections are summarized, and ordered, in Table~\ref{tab:coop_order}: the rate given in an upper row is greater than or equal to the one given in a lower row.
It is assumed that the cooperating nodes are close together such that $g>2$.
It can be observed that optimal power allocation contributes only marginal additional capacity gain over equal power allocation, while having full CSI (which allows the synchronization of the carriers of the transmitter and the relay) is essential to achieving any cooperative capacity gain.
Accordingly, in transmitter cooperation, homogeneous nodes with common battery and amplifier specifications may be employed to simplify network deployment, but synchronous-carrier should be considered necessary.
The nodes may synchronize through an external clock such as that provided by the Global Positioning System (GPS) \cite{hofmann-wellenhof06:gps_theo_prac}, or through exchanging timing reference signals \cite{elson02:time_sync_ref, sichitiu03:time_sync_sens}.
In \cite{jagannathan04:time_sync_errors} the performance penalty due to imperfect synchronization in cooperative systems is investigated.

\begin{table}
  \renewcommand{\arraystretch}{1.6}
  \centering
  \begin{tabular}{|l|l|}
    \hline
    \emph{Cooperation Scheme} & \emph{Rate}\\
    \hline\hline
    \Ct{1}, \Cr{1} & $\C\ns\left(\frac{2(g+1)}{g+2}\right)$\\
    \hline
    \Rt{1}, \Ct{2}, \Cr{3} & $\C\ns\left(\frac{2g}{g+1}\right)$\\
    \hline
    \Rt{2} & $\C\ns\left(\frac{2(g-1)}{g}\right)$\\
    \hline
    \Rpr\ & $\C\ns\left(\frac{2g(\sqrt{g-1}-1)(g-1-\sqrt{g-1})}
      {\sqrt{g-1}(g-2)^2}\right)$\\
    \hline
    \Rr{1}, \Rr{3} & ${\displaystyle\max_{0\leq\alpha\leq1}}
    \C\ns\left(\frac{\alpha(1-\alpha)g}{(1-\alpha)g+2\alpha+1/P}
      +\alpha\right)$\\
    \hline
    \Cn, \Cr{2}, \Ct{3}, \Rt{3}, \Ct{4}, \Rt{4}, \Cr{4} & $\C(1)$\\
    \hline
    \Rr{2}, \Rr{4} & $\C\ns\left(\frac{g}{2(g+2+2/P)}+\frac{1}{2}\right)$\\
    \hline
  \end{tabular}
  \caption{Cooperation rates comparison}
  \label{tab:coop_order}
\end{table}

On the other hand, in receiver cooperation, the compress-and-forward scheme does not require CSIT, but optimal power allocation is crucial in attaining cooperative capacity gain.
When channel phase information is not utilized at the transmitter (i.e., $\rho=0$), as noted in \cite{host-madsen06:coop_bounds}, carrier-level synchronization is not required between the relay and the transmitter; network deployment complexity is thus significantly reduced.
It is important, however, for the nodes to be capable of transmitting at variable power levels, and for the network power allocation to be optimized among the cooperating nodes.

\section{Fast Rayleigh Fading}
\label{sec:ray_fading_ch}

In this section we assume the channels in the relay network experience Rayleigh fading where the channel gain $h_i$ is iid ZMCSCG with unit variance, with the corresponding power gain being iid exponential with unit mean: $\gamma_i\triangleq\asq{h_i}\sim\exp(1)$, $i=1,2,3$.
We assume a fast-fading environment where the receivers have CSI but the transmitters know only CDI\@.
Furthermore, we consider the high SNR regime: when evaluating the capacity expressions, we make use of the high SNR ($P\gg1$) approximation $\log(1+xP) \cong \log(xP)$, in the sense that $\lim_{P\rightarrow\infty}\log(1+xP)-\log(xP)=0$, where $x>0$ is a given scalar.
We consider two cases of power allocation among the cooperating nodes: i) equal power allocation, and ii) optimal power allocation.
In each case we evaluate the cooperation cut-set bounds and achievable rates, and capacity gain from transmitter and receiver cooperation are compared to determine the best cooperation strategy based on the given power allocation assumption.

\subsection{Transmitter Cooperation}

The transmitters know only the fading distribution but do not have CSI\@.
Without CSIT, \cite{kramer05:coop_cap_relay, host-madsen05:power_relay} show that it is optimal to have the transmitter and the relay send uncorrelated signals in evaluating the cut-set bound and the decode-and-forward rate.
Suppose that the transmitter is operating under an average power constraint $\alpha P$, $0\leq\alpha\leq 1$, and
the relay under constraint $(1-\alpha)P$.
Then for the transmitter cooperation configuration depicted in Fig.~\ref{fig:txcoop}, the ergodic capacity cut-set bound is given by
\begin{align}
  \label{eq:fCt}
  \fCt{} = \min\Bigl\{
  \E\bigl[\,\C\bigl(\alpha(g\gamma_2+\gamma_1)\bigr)\bigr],\,
  \E\bigl[\,\C\bigl(\alpha\gamma_1 + (1-\alpha)\gamma_3\bigr)\bigr]
  \Bigr\}.
\end{align}
The parameter $\alpha$ is set to $1/2$ under equal power allocation in Section~\ref{sec:eq_pow}, but will need to be optimized when optimal power allocation is considered in Section~\ref{sec:opt_pow}\@.
Note that as argued in \cite{host-madsen05:power_relay}, since we assume each transmission block is long enough to undergo different fading realizations, the expectation is taken on each term inside the $\min\{\cdot\}$, instead of the entire expression.
The decode-and-forward transmitter cooperation rate is given by
\begin{align}
  \label{eq:fRt}
  \fRt{} = \min\Bigl\{
  \E\bigl[\,\C\bigl(\alpha g\gamma_2\bigr)\bigr],\,
  \E\bigl[\,\C\bigl(\alpha\gamma_1 + (1-\alpha)\gamma_3\bigr)\bigr]
  \Bigr\},
\end{align}
where the power allocation parameter $\alpha$ carries similar interpretations as described above.

\subsection{Receiver Cooperation}

For the receiver cooperation configuration shown in Fig.~\ref{fig:rxcoop}, the cut-set bound is
\begin{align}
  \label{eq:fCr}
  \fCr{} = \min\Bigl\{
  \E\bigl[\,\C\bigl(\alpha(\gamma_2+\gamma_1)\bigr)\bigr],\,
  \E\bigl[\,\C\bigl(\alpha\gamma_1 + (1-\alpha)g\gamma_3\bigr)\bigr]
  \Bigr\}.
\end{align}
In the compress-and-forward receiver cooperation strategy, the relay sends a compressed version of its observed signal to the receiver. In a static channel, the compression is realized using Wyner-Ziv source coding \cite{wyner76:rate_dist_side_info}, which exploits the correlation between the received signal of the relay and that of the receiver.
However, as noted in \cite{host-madsen05:power_relay}, in a fading channel Wyner-Ziv compression cannot be applied directly since the joint distribution of the received signals is not iid. In particular, the correlation between the observed signals depends on the channel realization.

To obtain an achievable rate, we can have the relay compress the signal without taking advantage of the side information.
In this case, the compression scheme reduces to the standard rate distortion function for a complex Gaussian source.
In general, for independent (but not identically distributed) Gaussian random variables (RVs), rate distortion is achieved by optimally allocating information bits to the RVs by the reverse water-filling scheme \cite[Sec.~13.3.3]{cover91:eoit}, in which a Gaussian RV with large variance is described with a higher resolution.

However, for comparing transmitter and receiver cooperation rates, it suffices to consider a simple compression scheme where a constant number of information bits is used for each forwarded observation.
In particular, we will show that the rate of this simple receiver cooperation scheme outperforms transmitter cooperation in the case of optimal power allocation.
As described in (\ref{eq:rx_relay}), the relay observes the signal $y_1 = h_2 x + n_1$.
Having CSIR, the relay knows the realization of $h_2$, and it inverts the channel gain to produce a unit-variance iid ZMCSCG source $\tilde{y}_1 = y_1/(h_2\sqrt{\alpha P+1/\gamma_2})$.
Suppose the relay forwards $\tilde{y}_1$ to the receiver at rate $R_3$, then from the rate distortion function for an iid complex Gaussian source, the squared error distortion is given by $D=2^{-R_3}$.

To determine $R_3$, we note that if the receiver is decoding messages from the transmitter at a rate of
$\E\bigl[\,\C(\alpha\gamma_1)\bigr]$, then concurrently it is also able to decode from the relay at rate
$R_3 = \E\bigl[\,\C\bigl((1-\alpha)g\gamma_3/(\alpha\gamma_1P+1)\bigr)\bigr]$,
based on the multiple access channel (MAC) ergodic capacity region \cite{tse98:fading_mac_1} as achieved through successive decoding.
We assume that within the cooperation cluster the relay is able to share its CSIR with the receiver, so the receiver can perform maximal-ratio combining (MRC) to achieve the receiver cooperation rate:
\begin{align}
  \label{eq:fRr}
  \fRr{} =
  \E\bigl[\,\C\bigl(\alpha\gamma_1+\alpha\gamma_2/(1+\hat{N})\bigr)\bigr],
\end{align}
where $\hat{N}$ is the variance of the Gaussian compression noise given by
\begin{align}
  \label{eq:cprn_noise}
  \hat{N} =\frac{D}{1-D}(\alpha\gamma_2P+1).
\end{align}

\subsection{Equal power allocation}
\label{sec:eq_pow}

Under equal power allocation, we set $\alpha=1/2$. At high SNR the first term inside the $\min\{\cdot\}$ of the cut-set bound (\ref{eq:fCt}) evaluates to
\begin{align}
  \label{eq:fCt1_1}
  \E\bigl[\,\C\bigl((g\gamma_2+\gamma_1)/2\bigr)\bigr]
  = \log P + \frac{g\log g}{g-1}-\log e^\gamma -1,
\end{align}
when $g>1$; where $\gamma$ is Euler's constant: $\gamma\cong 0.577$.
Similarly, the second term inside the $\min\{\cdot\}$ of the cut-set bound is given by
\begin{align}
  \label{eq:fCt1_2}
  \E\bigl[\,\C(\gamma_1/2 + \gamma_3/2)\bigr]
  = \log P + \log e^{1-\gamma}-1.
\end{align}
Comparing the two terms inside the $\min\{\cdot\}$ expression of the cut-set bound, it can be observed that when $g>1$ (\ref{eq:fCt1_1}) always exceeds (\ref{eq:fCt1_2}).
Therefore, the cut-set bound evaluates to
\begin{align}
\fCt{1} &= \log P + \log e^{1-\gamma}-1,
\end{align}
when $g>1$. The superscript represents the power allocation assumption: 1)~equal power allocation; and 2)~optimal power allocation later considered in Section~\ref{sec:opt_pow}.

The transmitter cooperation decode-and-forward rate (\ref{eq:fRt}) under equal power allocation can be evaluated following similar steps.
In particular, when $g\geq e\cong 2.718$, the transmitter cooperation rate meets the cut-set bound:
\begin{align}
\fRt{1} &= \log P + \log e^{1-\gamma}-1.
\end{align}
Hence decode-and-forward is capacity-achieving for $g\geq e$ under Rayleigh fading channels with only CDI at the transmitters.

For receiver cooperation, when $g>1$, the corresponding cut-set bound resolves to
\begin{align}
\fCr{1} &= \log P + \log e^{1-\gamma}-1,
\end{align}
which is identical to the transmitter cooperation cut-set bound.
However, there is no simple analytic expression for the compress-and-forward receiver cooperation rate (\ref{eq:fRr}). In the numerical example the compress-and-forward rate $\fRr{1}$ is evaluated using Monte Carlo simulation over 1000 instances of channel realizations.

\begin{figure}
  \centering
  \includegraphics*[width=8cm]{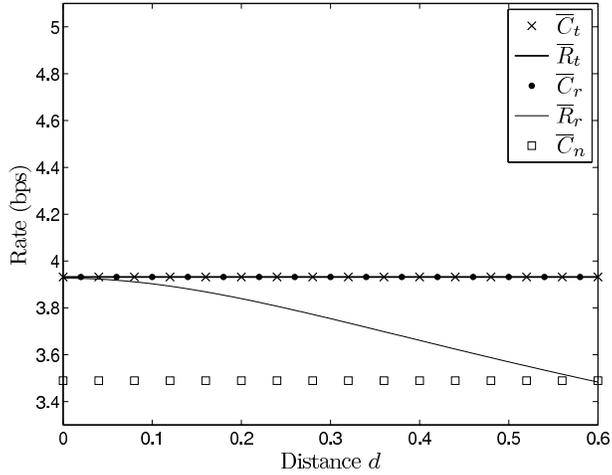}
  \caption{Cooperation rates under equal power allocation in fast Rayleigh fading.}
  \label{fig:coop_fade_eq}
\end{figure}

The transmitter and receiver cooperation upper bounds and achievable rates are shown in Fig.~\ref{fig:coop_fade_eq}.
As a comparison, also shown is the non-cooperative ergodic capacity $\fCn$ of a single-user channel under the same network power constraint $P$\@.
At high SNR the non-cooperative capacity is given by
\begin{align}
\fCn & = \E\bigl[\,\C(\gamma_1)\bigr] = \log P - \log e^\gamma.
\end{align}
It can be observed that the transmitter cooperation capacity outperforms the compress-and-forward receiver cooperation rate, which in turn outperforms the non-cooperative capacity for small $d$.
Moreover, since the transmitter cooperation rate also meets the receiver cut-set bound, transmitter cooperation is at least as good as any receiver cooperation scheme can theoretically achieve.

\subsection{Optimal power allocation}
\label{sec:opt_pow}

Under optimal power allocation, the parameter $\alpha$ needs to be optimized. Applying the high SNR approximation,
the first term inside the $\min\{\cdot\}$ expression of the transmitter cooperation cut-set bound (\ref{eq:fCt}) is given by
\begin{align}
  \label{eq:fCt2_1}
  \E\bigl[\,\C\bigl(\alpha(g\gamma_2+\gamma_1)\bigr)\bigr]
  = \log P + \log\alpha + \frac{g \log g}{g-1} - \log e^\gamma,
\end{align}
while the second term evaluates to
\begin{align}
  \label{eq:fCt2_2}
  &\E\bigl[\,\C\bigl(\alpha\gamma_1 + (1-\alpha)\gamma_3\bigr)\bigr] =
  \begin{cases}
    \log P +
    \frac{\alpha\log\alpha-(1-\alpha)\log(1-\alpha)}{2\alpha-1} - \log
    e^\gamma & \text{if $\alpha\neq 1/2$,}\\
    \log P + \log e^{1-\gamma} - 1 & \text{if $\alpha = 1/2$}.
  \end{cases}
\end{align}
Note that (\ref{eq:fCt2_2}) is concave in $\alpha$ and its maximum is achieved at $\alpha = 1/2$.
Further, if $g\geq 1$, (\ref{eq:fCt2_1}) exceeds (\ref{eq:fCt2_2}) at $\alpha=1/2$.
Hence when $g\geq 1$, the optimal $\alpha^*$ is $1/2$, and the cut-set bound is the same as that under equal power allocation:
\begin{align}
\fCt{2} &= \log P + \log e^{1-\gamma}-1.
\end{align}
Similar steps can be used to calculate the decode-and-forward transmitter cooperation rate.
If $g\geq e \cong 2.718$, it is found that equal power allocation is optimal ($\alpha^*=1/2$), and the decode-and-forward achievable rate meets its cut-set bound:
\begin{align}
\fRt{2} &= \log P + \log e^{1-\gamma}-1.
\end{align}
Hence, under Rayleigh fading with only CDI at the transmitter, it follows that when the relay is close to the transmitter, decode-and-forward with equal power allocation is optimal and capacity-achieving.

For receiver cooperation, at high SNR the first term inside the $\min\{\cdot\}$ expression of the cut-set bound
(\ref{eq:fCr}) is given by
\begin{align}
  \label{eq:fCr2_1}
  \E\bigl[\,\C\bigl(\alpha(\gamma_2+\gamma_1)\bigr)\bigr]
  = \log P + \log\alpha + \log e^{1-\gamma},
\end{align}
while the second term evaluates to
\begin{align}
  \label{eq:fCr2_2}
  \E\bigl[\,\C\bigl(\alpha\gamma_1 + (1-\alpha)g\gamma_3\bigr)\bigr]
  = \log P +
  \frac{g(1-\alpha)\log(g(1-\alpha))-\alpha\log\alpha}
  {g(1-\alpha)-\alpha}.
\end{align}
We note that when $g\geq 1$, the optimal power allocation $\alpha^*$ is achieved by equating (\ref{eq:fCr2_1}) and (\ref{eq:fCr2_2}), which can be solved numerically.
The solution to $\alpha^*$ can then be used to compute the cut-set bound $\fCr{2}$.
For the compress-and-forward receiver cooperation rate (\ref{eq:fRr}), it is evaluated using Monte Carlo simulation over 1000 instances of channel realizations.
To ease computational complexity, the optimization is relaxed to find the maximum compress-and-forward rate $\fRr{2}$ with the power allocation parameter $\alpha$ varying in discrete steps of 0.01.

\begin{figure}
  \centering
  \includegraphics*[width=8cm]{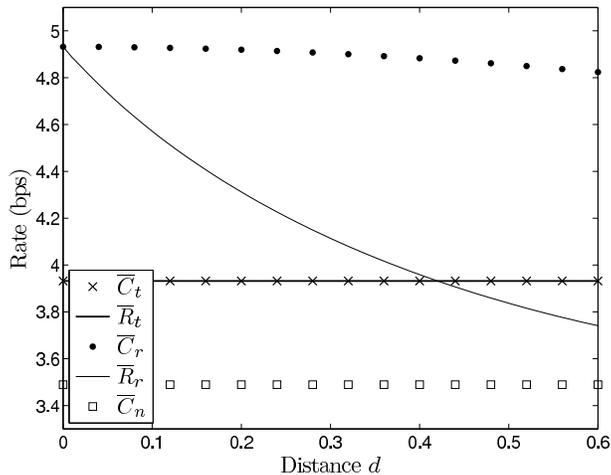}
  \caption{Cooperation rates under optimal power allocation in fast Rayleigh fading.}
  \label{fig:coop_fade_opt}
\end{figure}

The transmitter and receiver upper bounds and achievable rates are shown in Fig.~\ref{fig:coop_fade_opt}.
Both transmitter and receiver cooperation offer capacity gain over the non-cooperative scheme.
However, in contrast to the case of equal power allocation, in this case the compress-and-forward receiver cooperation rate is higher than the transmitter cooperation cut-set bound for small $d$.
Consequently, similar to the findings under quasi-static channels, we conclude that receiver cooperation is the superior strategy when the system is under optimal power allocation.

\subsection{Effects of Fast Rayleigh Fading}

\begin{figure}
  \centering
  \includegraphics*[width=8cm]{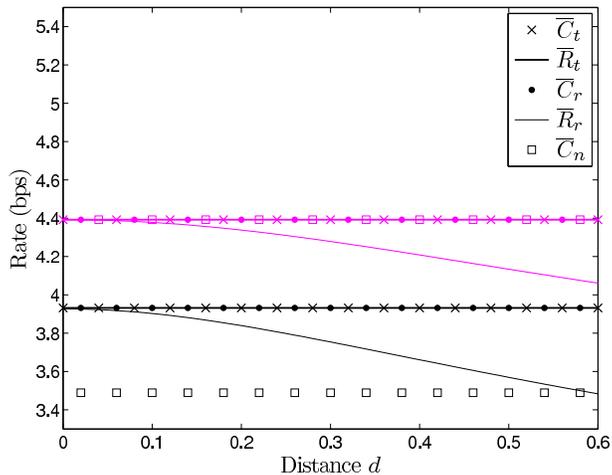}
  \caption{Effects of fading under equal power allocation. In the darker shade are the cooperation rates under fast Rayleigh fading, while the lighter shade represents the rates under quasi-static phase fading.}
  \label{fig:eff_fade_eq}
\end{figure}

\begin{figure}
  \centering
  \includegraphics*[width=8cm]{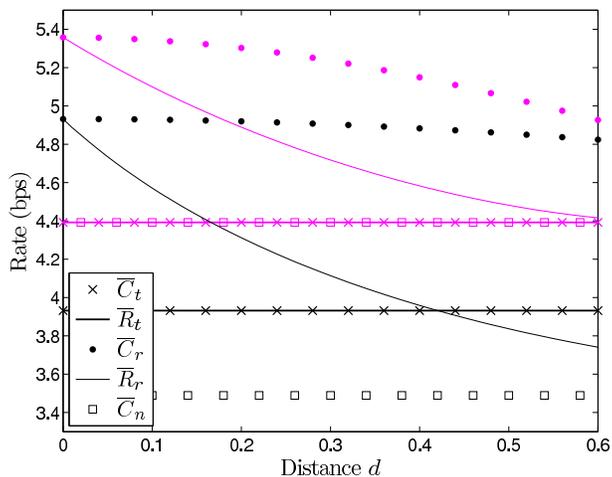}
  \caption{Effects of fading under optimal power allocation. The darker shade represents fast Rayleigh fading, and the lighter shade represents quasi-static phase fading.}
  \label{fig:eff_fade_opt}
\end{figure}

In this section, we contrast the cooperative capacity improvement under quasi-static phase fading in Cases~3 and 4 in Section~\ref{sec:coop_rates} versus that under fast Rayleigh fading in Sections~\ref{sec:eq_pow} and \ref{sec:opt_pow} to characterize the effects of fast fading in channel magnitudes on cooperation.
We assume the nodes have only CSIR (but all nodes have CDI), and examine the cooperation rates under the cases of equal power allocation and optimal power allocation.
Under equal power allocation, the cooperation rates are plotted in Fig.~\ref{fig:eff_fade_eq}, while the rates under optimal power allocation are shown in Fig.~\ref{fig:eff_fade_opt}.
As expected, fading decreases capacity.
However, it can be observed that fading in channel magnitudes diminishes the non-cooperative capacity more severely than it does the cooperation rates.
In particular, under equal power allocation, cooperation provides no capacity gain in the quasi-static channels; whereas cooperation improves capacity under fast Rayleigh fading.
Similarly, under optimal power allocation, the capacity gain under fast Rayleigh fading is higher than that in the quasi-static channels.
The resilience of a cooperative system to fading comes from transmitter and receiver diversity analogous to that in a multiple-antenna channel.
Without CSIT, having multiple transmit antennas offers no capacity improvement in a static environment \cite[Sec.~4.4]{paulraj03:intro_st_wcom}.
However, in a fading environment, even in the absence of CSIT, multiple transmit antennas mitigate the effects of fading through diversity, which leads to a higher capacity than that of a corresponding single-input single-output (SISO) channel.

On the other hand, when we have fading in the channel magnitudes the cooperating nodes need to be within a closer range in order for the decode-and-forward cooperation scheme to achieve capacity
(decode-and-forward requires $g\geq 2.718$ to achieve capacity under fast Rayleigh fading vs.\ $g\geq 1$ in the quasi-static channels).
Moreover, the transmitter cooperation capacity is more sensitive to power allocation under fast Rayleigh fading:
Any deviation from the optimal allocation $\alpha^*=1/2$ diminishes capacity in the fast Rayleigh fading channels; whereas in the quasi-static channels, the optimal allocation $\alpha^*$ can be any value in the range $[1/g, 1]$ to be capacity-achieving.

\subsection{Cooperative Capacity Gain Necessary Conditions}

\begin{figure}
  \centerline{%
    \subfigure[Transmitter cluster]{%
      \psfrag{h1}[][]{\small$h_1$}
      \psfrag{hM}[][]{\small$h_M$}
      \includegraphics{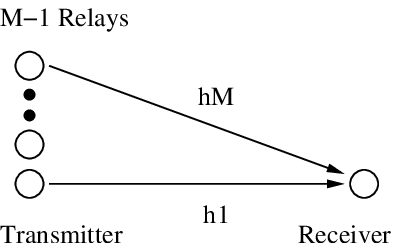}
      \label{fig:m_tx}
    }
    \hfil
    \subfigure[Receiver cluster]{%
      \psfrag{h1}[][]{\small$h_1$}
      \psfrag{hM}[][]{\small$h_M$}
      \includegraphics{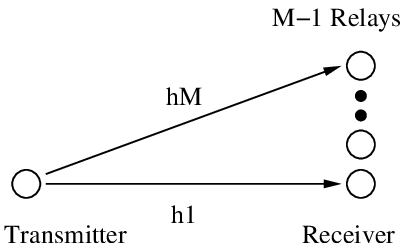}
      \label{fig:m_rx}
    }
  }
  \caption{Cooperation in transmitter and receiver clusters.}
  \label{fig:coop_network}
\end{figure}

In the previous sections we consider small cooperative networks comprising three nodes;
in this section we generalize the necessary conditions for cooperative capacity gain for a transmitter or receiver cluster with a large number of cooperating nodes.
As depicted in Fig.~\ref{fig:coop_network}, suppose there are $M$ single-antenna cooperating nodes in the transmitter cluster or the receiver cluster.
When we have quasi-static phase fading channels, $h_i=e^{j\theta_i}$, $\theta_i\sim U[0,2\pi]$ iid, $i=1\ldots M$; whereas under fast Rayleigh fading, the channel gains $h_i$s are iid unit-variance ZMCSCGs.
Again, there is a network power constraint of $P$ on the system.

First consider the quasi-static phase fading channels.
For the transmitter cluster shown in Fig.~\ref{fig:m_tx}, its capacity under optimal power allocation is upper-bounded by that of an $M$-antenna multiple-input single-output (MISO) channel.
Without CSIT (i.e., $\theta_i$'s unknown to the transmitter), the MISO channel has the same capacity $\C(1)$ as that of a SISO channel \cite{paulraj03:intro_st_wcom}.
For the receiver cluster shown in Fig.~\ref{fig:m_rx}, if equal power allocation is imposed, then the transmitter and $M-1$ relays are each under power constraint $P/M$.
Likewise, the capacity of an $M$-antenna single-input multiple-output (SIMO) channel serves as an upper bound
to the capacity of the receiver cooperation cluster.
With maximal-ratio combining at the receiver, the SIMO capacity is again $\C(1)$.
Next consider the similar upper bounds under fast Rayleigh fading, when the number of cooperating nodes $M$ is large.
From the strong law of large numbers, the multiple antennas remove the effects of fading, and both the MISO
and SIMO channels achieve the same capacity of $\C(1)$, asymptotically in $M$.

Comparing the upper bounds to the non-cooperative capacity $\Cn =\C(1)$ under quasi-static phase fading, and $\fCn \cong \C(1) - 0.833$ under
fast Rayleigh fading, we observe that transmitter cooperation without CSIT, or receiver cooperation under equal power allocation, provides no capacity gain under quasi-static phase fading, and at most a constant capacity gain that fails to grow with $M$ under a fast fading environment.
Therefore, to achieve the $\Theta(\log M)$ capacity scaling as reported in \cite{gastpar05:large_relay} in a large Gaussian relay network of $M$ nodes, CSIT is necessary for transmitter cooperation, and inhomogeneous power allocation is necessary for receiver cooperation.

\section{Conclusion}
\label{sec:conclu}

We have studied the capacity improvement from transmitter and receiver cooperation when the cooperating nodes form a cluster in a relay network in quasi-static phase fading channels and fast Rayleigh fading channels.
It was shown that electing the proper cooperation strategy based on the operating environment is a key factor in realizing the benefits of cooperation in a wireless network.
Under quasi-static phase fading, when full CSI is available, transmitter cooperation is the preferable strategy.
On the other hand, when CSIT is not available but power can be optimally allocated among the cooperating nodes, the superior strategy is receiver cooperation.
When the system is under equal power allocation with CSIR only, cooperation offers no capacity improvement over a non-cooperative single-transmitter single-receiver channel under the same network power constraint.
Similar conclusions follow in a fast Rayleigh fading environment.
Under fast Rayleigh fading in the high SNR regime, it was shown that under equal power allocation, the decode-and-forward transmitter cooperation scheme is capacity-achieving and is superior to receiver cooperation.
On the other hand, under optimal power allocation, the compress-and-forward receiver cooperation scheme outperforms transmitter cooperation.

In addition, by contrasting the cooperative capacity gain under fast Rayleigh fading to that under quasi-static phase fading, we have shown that cooperative systems provide resilience to fading in channel magnitudes.
However, under fast Rayleigh fading, capacity becomes more sensitive to power allocation, and the cooperating nodes need to be closer together in order for the decode-and-forward transmitter cooperation scheme to achieve capacity.
Finally, necessary conditions on cooperative capacity gain were investigated.
It was shown that in a large cluster of $M$ cooperating nodes, transmitter cooperation without CSIT or receiver cooperation under equal power allocation provides no capacity gain under quasi-static phase fading, and at most a constant capacity gain that fails to grow with $M$ in a fast fading environment.

\bibliographystyle{IEEEtran}
\bibliography{IEEEabrv,wrlscomm}


\end{document}